\documentstyle[aps,12pt,prd]{revtex}

\begin{document}
\title{Defect Formation in First Order Phase Transitions with Damping }
\author{Antonio Ferrera \thanks{%
e-mail: aferrera@pinar2.csic.es}}
\address{Instituto de Matematica Aplicada y Fisica Fundamental,\\
Consejo Superior de Investigaciones Cientificas,\\
Serrano 121, Madrid}
\date{20/12/96}
\maketitle

\begin{abstract}
Within the context of first order phase transitions in the early universe,
we study the influence of a coupling between the (global U(1)) scalar
driving the transition and the rest of the matter content of the theory. The
effect of the coupling on the scalar is simulated by introducing a damping
term in its equations of motion, as suggested by recent results in the
electroweak phase transition. After a preceeding paper, in which we studied
the influence that this coupling has in the dynamics of bubble collisions 
and topological defect formation, we proceed in this paper to
quantify the impact of this new effects on the probability of defect
creation per nucleating bubble.
\end{abstract}


\tighten
\preprint{\vbox{
\hbox{CAT-95/05}}}

\section{Introduction}

According to the standard model and its extensions, symmetry breaking phase
transitions are expected to occur in the early universe. The mechanism by
which these transitions may take place could be either by the formation of
bubbles of the new phase within the old one or by spinodal decomposition
(i.e., either by a first order phase transition or by a second order one).
In the particular case of the electroweak phase transition, for instance,
common opinion inclines more towards the first of the two possibilities. As
this scenario would have it, bubbles of the new phase nucleated within the
old one (the nucleation process being described by instanton methods as far
as the WKB approximation remains valid \cite{c77}), and subsequently
expanded and collided with each other until they occupied all of the
available volume at the time at which the transition was completed. In the
process of bubble collision though, the possibility arises that regions of
the old phase become trapped within the new one, giving birth to
topologically stable localized energy concentrations known as {\it %
topological defects } (for recent reviews see refs. \cite{td-rev} ), much in
the same way in which these structures are known to appear in condensed
matter phase transitions. From a theoretical point of view, topological
defects will appear whenever a symmetry group $G$ is spontaneously broken to
a smaller group $H$ such that the resulting vacuum manifold $M=G/H$ has a
non-trivial topology: cosmic strings for instance (vortices in two space
dimensions) will form whenever the first homotopy group of $M$ is
non-trivial, i.e., $\pi _1(M)\neq 1.$

To see how this could happen in detail, let's consider the Lagrangian

\begin{equation}
{\cal L}=\frac 12\partial _\mu \Phi \partial ^\mu \Phi ^{*}-V(\Phi )
\label{lag}
\end{equation}
for a complex scalar field $\Phi $. Let us assume that $V$ is of the type $V=%
\frac \lambda 4(|\Phi |^2-\eta ^2)^2$, and that its parameters are functions
of the temperature such that at high temperatures $\Phi =0$ is the only
minimum of $V$, while at zero temperature all the $|\Phi |=\eta $ states
correspond to different degenerate minima. Then the structure of the vacuum
manifold will be that of $S^1$. $\pi _1(S^1)\neq 1$ however, and thus we can
form non-contractible loops in the vacuum manifold: the model admits cosmic
string solutions.

The way in which these configurations would actually form during a phase
transition is via the Kibble mechanism \cite{k76}. In the context of a first
order transition the basic idea behind it is that bubbles are nucleated with
random phases of the field, and that when two regions in which the phase
takes different values encounter each other the phase should interpolate
between this two regions following a geodesic path in the vacuum manifold
--the so-called geodesic rule. A possible scenario for vortex formation in
two space dimensions would then look like this: three bubbles with
respective phases of $0,$ $2\pi /3,$ and $4\pi /3$ collide simultaneously.
Thus, if we walk from the first bubble to the second , then from the second
to the third, and finally back from the third to the first one again, the
phase will have wounded up by $2\pi $ in our path, having traversed the
whole of the vacuum manifold once along the way. Continuity of the field
everywhere inside the region contained by our path demands then that the
field be zero at some point inside of it, namely the vortex core. In the
limit in which the bubbles extend to infinity, outwards from the center of
collision, removal of this vortex would cost us an infinite amount of
energy, since it would involve unwinding the field configuration over an
infinite volume. The vortex is thus said to be topologically stable. In
three space dimensions, the resulting object would obviously be a string,
rather than a vortex.

Clearly, there are other ways in which strings could be formed. Collisions
of more than three bubbles could also lead to string formation, or, for
instance, two of the bubbles could hit each other first, with the third one
hitting only at some later time while the phase is still equilibrating
within the other two. This event in particular will be far more likely than
a simultaneous three way (or higher order) collision, and it is probably the
dominating process by which strings are formed (especially if nucleation
probabilities are as low as required for WKB methods to be valid). When two
bubbles collide, the phase will try to reach a homogeneous distribution
within the single true vacuum cavity formed after the collision. However, in
the absence of coupling between the scalar and the rest of the matter this
process is never completed --essentially because the velocity at which the
phase propagates inside the bubbles is the same as that with which the
bubble walls expand--, and thus this does not substantially modify the
picture of defect formation described above.

Bubble collisions have been studied by a number of groups, most notably by
Hawking {\it et al.} \cite{hms82}, Hindmarsh {\it et al. } \cite{hdb94},
Srivastava \cite{s92}, Melfo and Perivolaropoulos \cite{mp95}, and more
recently by the author and Melfo \cite{fm96}. Only in this last work however
did the authors concern themselves with the interaction between the bubble
field and the surrounding plasma however, despite of the fact that from
current work in the electroweak phase transition we can expect that in many
cases such interaction will not be negligible (e.g., see Ref. \cite{ew}).
The basic underlying reason is relatively simple to understand: as the
bubble wall sweeps through an specific point, the Higgs field $\Phi $
acquires an expectation value, and the fields coupled to it acquire a mass.
Thus, particles with not enough energy to acquire their corresponding mass
inside the bubble will bounce off the wall (thus imparting negative momentum
to it), while the rest will get through. Obviously, the faster the wall
propagates the stronger this effect will be, since the momentum transfer in
each collision will be larger, and therefore a force proportional to the
velocity with which the wall sweeps through the plasma should appear. In the
overdamped regime then bubble walls will reach a terminal velocity in its
expansion, which by most accounts will probably not be relativistic. For
instance in \cite{ew} a value $v_{ter}\sim 0.1$ is predicted, although
higher estimates do exist in the literature (e.g., see \cite{mt95}). In this
situation one could expect the process of phase equilibration after two
bubbles collide --and thus of defect formation-- to be different from that
usually understood to take place when no dissipation is present and the
walls acquire relativistic speeds. Indeed, in 1994 Kibble \cite{kib94}
suggested that these differences could be important enough to result in a
lower value for the density of topological defects created during a full
phase transition in the damped scenario. To investigate in detail these
differences was the aim of ref. \cite{fm96}. In it, new physical effects
were found that modify the picture of how topological defects are formed
when viscosity plays an important role in the motion of the bubble walls.
The most salient feature of this new picture is perhaps the fact that, after
they collide, the two bubbles behave as a resonant cavity where the spatial
profile for the phase oscillates from its initial value to its inverse (that
is, with the phases inverted with respect to their initial spatial
distribution), then back to the initial profile, and so on. Therefore, an
immediate consequence of the existence of this oscillating state is that it
becomes possible, for the same set of three bubbles, to form a vortex, an
antivortex or no defect at all depending on the precise timing of the last
collision, in clear contrast to what occurs in the undamped scenario, where
the initial phases of the bubbles determine the type of defect that will be
formed. It was also seen there that these oscillations of the phase can
potentially last for a long time (up to $25R$ for $v_{ter}\sim 0.1$, where $%
R $ is the radii of the bubbles at collision time). As a consequence, the
two bubble system would have to remain in ``isolation'' for a very long time
before acquiring a homogeneous phase, a very unlikely situation.

The preceding arguments would seem to suggest that there will not be a
strong suppression of defect formation in these ``slow'' transitions,
although it is still rather unclear exactly how much will these events be
suppressed. The aim of this paper is thus to quantitatively investigate
precisely this question: what will be final impact that this new dynamic of
defect formation will have on the probability of forming defects?. In order
to do this we have performed a series of simulations of full phase
transitions for different values of the friction coefficient, $\gamma ,$
that phenomenologically models the coupling between the scalar driving the
transition and the rest of the matter, thus finding the behavior of the
density of nucleated defects per bubble as a function of $\gamma .$ The
paper is organized as follows: in section II we give a brief account of the
model and the results previously found in ref. \cite{fm96}, in section III
we present the results of the simulations, finally, the conclusions are
presented in section IV.

\section{2 and 3 bubble collisions in a damping environment}

Consider the Lagrangian (\ref{lag}) for a complex field $\Phi $. We will use
the same form of potential that was used in \cite{s92,mp95}, that is, 
\begin{equation}
V=\lambda \left[ \frac{|\Phi |^2}2\left( |\Phi |-\eta \right) ^2-\frac 
\epsilon 3\eta |\Phi |^3\right] .  \label{pot}
\end{equation}
This is just a quartic potential with a minimum at $|\Phi |=0$ (the false
vacuum), and a set of minima connected by a $U(1)$ transformation (true
vacuum) at $|\Phi |=\rho _{tv}\equiv \frac \eta 4(3+\epsilon +\sqrt{\left(
3+\epsilon \right) ^2-8})$, towards which the false vacuum will decay via
bubble nucleation. It is the dimensionless parameter $\epsilon $ that is
responsible for breaking the degeneracy between the true and the false vacua.

The equations of motion for this system are then 
\begin{equation}
\partial _\mu \partial ^\mu \Phi =-\partial V/\partial \Phi .  \label{em}
\end{equation}
For the potential (\ref{pot}), approximate solutions of (\ref{em}) exist for
small values of $\epsilon$, the so-called thin wall regime \cite{l83}, and
are of the form

\begin{equation}
|\Phi |=\frac{\rho _{tv}}2\left[ 1-\tanh \left( \frac{\sqrt{\lambda }\eta }2%
(\chi -R_0)\right) \right] ,  \label{thin}
\end{equation}
where $R_0$ is the bubble radius at nucleation time and $\chi ^2=|\stackrel{%
\rightarrow }{x}|^2-t^2$. The bubble then grows with increasingly fast speed
and its walls quickly reach velocities of order 1. We are interested however
in investigating a model with overdamped motion of the walls due to the
interaction with the surrounding plasma. In order to model this effect, we
will insert a frictional term for the modulus of the field in the equation
of motion, namely 
\begin{equation}
\partial _\mu \partial ^\mu \Phi +\gamma \stackrel{\cdot }{|\Phi |}%
e^{i\theta }=-\partial V/\partial \Phi ,  \label{damp-em}
\end{equation}
where $\stackrel{\cdot }{|\Phi |}\equiv $ $\partial |\Phi |/\partial
t,\theta $ is the phase of the field, and $\gamma $ stands for the friction
coefficient (which will as a matter of fact serve as parameter under which
we will hide our lack of knowledge about the detailed interaction between
the wall and the plasma). It can be shown (see \cite{fm96}) that this
equation does indeed posses a solution that shows the desired type of
overdamped motion for the bubble walls. In the thin wall limit, this
solution can be written as

\begin{equation}
\rho =\frac{\rho _{tv}}2\left[ 1-\tanh \left( \frac{\sqrt{\lambda }\eta }2%
\frac{(r-v_{ter}t-R_0)}{\sqrt{1-v_{ter}^2}}\right) \right] ,  \label{wall}
\end{equation}
which is simply a Lorentz-contracted moving domain wall with a velocity $%
v_{ter}$ of the form $v_{ter}\sim \epsilon \delta _m/\gamma \rho _{tv}^2$,
where $\delta _m$ is the bubble wall thickness.

In two bubble collisions, the phase will at first --before the collision
actually takes place-- interpolate between the values that it takes in each
bubble by means of a phase wall situated at the midpoint in between the
bubbles. This can be seen by taking the value of the modulus of the field $%
\rho $ far away from the bubbles to be given by the ansatz

\begin{eqnarray}
\Phi (bubble1+bubble2)\equiv \rho e^{i\theta }\simeq \Phi (bubble1)+\Phi
(bubble2)\equiv \rho _1e^{i\theta _1}+\rho _2e^{i\theta _2},  \label{apsol}
\end{eqnarray}
together with the asymptotic forms for $\rho _{1,}\rho _2$ (in a 1
dimensional approximation where the bubble centers are situated at $\pm x_0$)

\begin{eqnarray}
\rho _1 &\simeq &\rho _{tv}e^{(x+vt-x_0)/\delta _m},  \label{asympt} \\
\rho _2 &\simeq &\rho _{tv}e^{-(x-vt+x_0)/\delta _m},  \nonumber
\end{eqnarray}
and plugging the value thus obtained for $\rho $ into the equation of motion
for the phase $\theta .$ Note that we are assuming that (\ref{apsol}) yields
only a correct approximation for $\rho $, and not for $\theta $, since the
motion of the two bubbles as they approach each other could in principle
generate phase waves at the midpoint between them. The resulting equation
for $\theta $ will depend on the phase difference between the two bubbles $%
\Delta \theta $. The most interesting value for this difference is however $%
\Delta \theta =\pi /2$, since for $\Delta \theta =0$ there is no dynamics to
the phase, and for $\Delta \theta =\pi $ the phase is undefined at the
midpoint between the bubbles. A solution interpolating from say $\theta _1=0$
at $x\rightarrow -\infty $ to $\theta _2=\pi /2$ at $x\rightarrow +\infty $
is

\begin{equation}
\theta =\frac 12\arcsin (\tanh (\frac{2x}{\delta _m}))+\frac \pi 4,
\label{pwall}
\end{equation}
which clearly shows the structure of a phase wall placed at the origin and
of width closely related to the bubble wall's width $\delta _m$. Time
dependent solutions to the equation for $\theta $ representing travelling
waves do exist, but they die out in time scales of the order of $\delta
_m/v_{ter}$.

At the moment of collision, and while the bubble walls merge with one
another, this phase wall will thicken up to a value of {\it at least} twice
the thickness of the bubble walls (since the time the walls take to complete
the merging is $t_{merging}\sim \delta _m/v_{ter}$ and the thickening occurs
at both sides of the phase wall). After the two bubbles have merged then,
the phase is free to travel in the resulting true vacuum cavity since it is
a Goldstone boson. The equation of motion for the phase is simply a wave
equation, of which approximate solutions can be found by assuming SO(1,2)
invariance in the $(t,x,y)$ subspace (the bubbles nucleate at $(0,0,0,\pm R)$%
). This symmetry is exact in the undamped case, but it still adequately
describes the behavior of the phase in the damped scenario, especially in
the bubble interior, far from the walls. With the initial conditions

\begin{equation}
\theta |_{\tau =0}=\theta _0\varepsilon (z)\;,\;\;\;\;\partial _\tau \theta
|_{\tau =0}=0,  \label{inicond}
\end{equation}
where $\tau ^2=\ t^2-x^2-y^2,$ the solution to the equation of motion is

\begin{equation}
\theta =\left\{ 
\begin{array}{c}
\theta _0\;\qquad \text{ for\qquad }z>\sqrt{t^2-x^2-y^2}, \\ 
\theta _0\frac z{\sqrt{t^2-x^2-y^2}}\text{\quad for \ \ }|\,z\,|\leq \sqrt{%
t^2-x^2-y^2}, \\ 
-\theta _0\;\qquad \text{ for\qquad }-z>\sqrt{t^2-x^2-y^2}.
\end{array}
\right.  \label{prowaves}
\end{equation}
At some point after these waves start to propagate into each bubble, they
will inevitably catch up with the bubble walls since they now move with a
speed less than 1 due to the viscosity. It is possible to perform an
analysis of the asymptotic fate suffered by an incoming plane wave of
frequency $\omega $ which hits the bubble wall; the results show that the
reflection coefficient $R$ will be given by (see \cite{fm96})

\begin{equation}
R=\left\{ 
\begin{array}{c}
1\qquad \qquad \qquad \qquad \;\;\text{for \quad }|\omega |\delta _m<\beta
\left( 1+v_{ter}\right) , \\ 
\frac{\sinh ^2\left( \pi \left( \omega \delta _m-\,\stackrel{\symbol{126}}{%
\omega }\right) \right) }{\sinh ^2\left( \pi \left( \omega \delta _m+\,%
\stackrel{\symbol{126}}{\omega }\right) \right) }\qquad \text{for \quad }%
|\omega |\delta _m>\beta \left( 1+v_{ter}\right) ,
\end{array}
\right.  \label{refcoef}
\end{equation}
where $\beta \equiv 1/\sqrt{1-v_{ter}^2}$, and $\stackrel{\symbol{126}}{%
\omega }=\sqrt{\omega ^2\delta _m-1}$ . The appearance of the $\beta $
factor and the $v_{ter\text{ }}$summand in the conditions of (\ref{refcoef})
is obviously due to the fact that the incoming plane wave is colliding
against a moving wall. In the rest frame of the wall the condition for $R=1$
reads $|\omega ^{\prime }|\delta _m<1$, where $\omega ^{\prime }$ is the
frequency of the incoming wave in that frame. Thus, if the terminal velocity
of the wall is non-relativistic (\ref{refcoef}) basically tells us that the
incoming wave will be totally reflected{\em \ }by the wall if its wavelength
is larger than the wall thickness, and partially reflected and partially
transmitted if its wavelength is shorter than that --with the reflected part
tending to zero as the wavelength diminishes. Remember however that at the
moment of the bubble collision, during the merging of the walls, the
thickness of the phase wall grew up to a value of the order of twice the
bubble wall thickness, and more likely larger than that. It seems clear then
that all of the Fourier components of the phase wave will have wavelengths
that will fall into the total reflection regime, and thus the whole phase
wave itself will simply be reflected by the wall. Let us imagine for the
sake of clarity that, at collision time, bubble 1 had a phase $\theta _1$
and bubble two $\theta _2>\theta _1$, with $\theta _2-\theta _1=\Delta
\theta $. If the shape of the phase wall at the completion of the collision
was $f(x)$ then, after it, we will have phase waves with shape $%
f(x-t)/2,\,\,f(x+t)/2$ propagating into each bubble, carrying a phase
difference $-\Delta \theta /2$ into bubble 2 and $\Delta \theta /2$ into
bubble 1. After these waves have bounced off the bubble walls and propagated
back into the interior again, the phase of each bubble will however be, for
bubble 2, $\theta _2-2\Delta \theta /2=\theta _1$, and for bubble 1, $\theta
_1+2\Delta \theta /2=\theta _2$. The phases of the bubbles will thus have
switched. The whole process is depicted in Figure 1, where the referred
sequence has been plotted from a simulation. In Fig. 1a, the walls of the
two bubbles are just about to finish their merging (continuous line), and
the shape of the phase wall at that time is shown (dashed line). The bubble
to the right plays the role of bubble 2 above, having $\theta _2>\theta _1.$
The following pictures show how the two phase waves propagate into the
bubbles (Fig. 1b), and back after bouncing (Fig. 1c). As expected, after
reflection each phase wave still carries a phase of $\pm \Delta \theta /2,$
for bubbles 1 and 2 respectively. Finally, in Fig. 1d the two phase waves
meet again. The phase polarity of the system has been completely inverted.

The next relevant question to ask is how long will it take until the phase
equilibrates. The only mechanism through which the phase waves loose energy
is the Doppler shifting of their frequency due to the fact that they bounce
off a moving wall. A back of the envelope calculation shows that, in 1+1
dimensions and for a terminal velocity of the bubble walls of 0.1, the phase
waves will reach a wavelength of the order of the radius of the single true
vacuum cavity ${\cal R}$ after completing 5 oscillations, that is, after an
approximate time of 25$R$, where $R$ is the radius of each bubble at
collision time.

In 2+1 dimensions the process develops along the same qualitative lines, as
we can see in Figs.2 and 3, where the contour lines of the phase (in units
of $\pi $) are plotted in continuous lines, while the bubble walls appear as
dashed lines. In two spatial dimensions, at any point that the phase wave
meets the wall, its propagation vector will have one component in the
direction normal to the wall and another one tangential to it. Only the
component in the normal direction will see the wall and bounce off it
however, while the other one will continue to propagate freely. Thus,
although in general the interaction between the wall and the phase wave will
be a complicated superposition of these two processes, we can expect that
after some time, in the region close the walls the phase will predominantly
propagate tangentially to them, the rest of it having bounced towards the
center. The developing of this process is depicted in Fig. 2, and its final
stage is clearly visible in Fig. 3(a). At that point virtually all the phase
propagating normal to the wall has bounced towards the center, and
tangential propagation dominates close to the wall. Now we only have to wait
for the central region of the phase, propagating along the $z$ axis, to get
to the end of the bubble and bounce off the wall. In Figures 3(b), (c) and
(d) we see how this takes place. While the area of the phase fronts
propagating along the $z$ axis collides head on with the end of the bubbles
and bounces as expected, the tangential components cross each other at that
region. Note how while the bouncing and crossing is taking place the phase
has a virtually homogeneous distribution inside the bubbles (Figs 3(b) and
(c)). After this, the combination of these two phenomena brings about an
inversion of the phase similar to that found in 1+1 dimensions, as can be
clearly seen in Fig. 3(f). The main consequence that this process will have
on the dynamics of vortex formation is now obvious: it will become possible,
for the same set of three bubbles, to form a vortex, no defect at all, or an
antivortex depending on the precise timing of the last collision (i.e., on
the state of the resonant cavity at the moment at which the collision with
the third bubble takes place).

The lifetime of the oscillating state is also very large in 2+1 dimensions,
and we did observe some vortex nucleation events even after the phase had
completed the 5 oscillations that we estimated for the relaxation time in
1+1 dimensions. Since this state is so long-lived, the two bubble system
will not reach a homogeneous phase distribution until very late into the
phase transition, if it ever does. Therefore, a full quantification of its
effects on the probability of nucleating vortices needs to take into account
multiple bubble collisions. This is the goal of the next section.

\section{Many bubble collisions: vortex formation probability as a function
of the friction coefficient}

In order to be able to quantify the impact of the type of frictional
coupling that we are considering on vortex formation probabilities we must
then carry out simulations involving many bubble collisions. To explore the
dependency of the vortex formation probabilities on the value of the
friction coefficient (or what is the same, on the value of the terminal
velocity for the bubble expansion), we have performed a series of
simulations for $\gamma $ ranging from $0.25$ (close to the undamped
situation where $\gamma =0$) up to a value of $\gamma =100$ (corresponding
to a terminal velocity for the bubble walls expansion of $v_{ter}\sim 10^{-2}
$). Our simulations differ from the standard numerical simulations of defect
production \cite{VacVil84} in some important aspects. In them, relative
phases are assigned at random to sites on a lattice corresponding to the
centres of causally disconnected regions of true vacuum (either bubbles in a
first-order or domains in a second-order transition). Between these sites
the phase is taken to vary along the shortest path on the vacuum manifold
--the geodesic rule. Defects are then formed wherever this geodesic
interpolation between sites generates a topologically nontrivial path in the
vacuum manifold. For a first-order transition this formalism corresponds to
true vacuum bubbles nucleating simultaneously, equidistant from all their
nearest neighbors. Consequently all collisions between neighboring bubbles
occur simultaneously, and the associated phase differences are simply given
by the differences in the initial assigned phases. We on the other hand have
used a leap-frog method where the discretization of space and time was such
that the grid and time step sizes were several times smaller than the bubble
wall thickness, and included the 'exact' (discretized) field dynamics in the
simulation. This of course was needed in order to include the effects of
phase bouncing and oscillations described in the preceding section. To take
full advantage of this approach then, bubbles were not nucleated equidistant
and simultaneously. We have followed \cite{vvkb} in nucleating the bubbles
at random points in space and time during the course of the simulation. The
algorithm thus goes as follows:

\begin{enumerate}
\item  generate a population of time ordered bubble nucleation events
distributed randomly within some finite volume of 2+1 dimensional
space-time, assigning a random phase to each bubble.

\item  start with an initial state for which the field is in the false
vacuum in all the simulation volume.

\item  at the beginning of each time step check whether there are any
bubbles to be nucleated at that time:

\begin{enumerate}
\item  if there are any bubble nucleation events, nucleate only those
bubbles that would fall entirely in a false vacuum region and discard the
rest (i.e., avoid superimposing new bubbles on regions that are already in
the true vacuum).
\end{enumerate}

\item  evolve the resulting field configuration to the next time step,
following the field equation (\ref{damp-em}), using a leap-frog method.
\end{enumerate}

In order to generate the bubble nucleation events, we first fix their number
--multiplying a sufficiently low but otherwise arbitrarily chosen nucleation
rate by the simulation volume--, then choose at random the space-time points
at which they take place within the simulation volume. Since we only
nucleate those bubbles that fall within the false vacuum and discard the
rest, at later times into the transition it will become increasingly
difficult for the new bubbles to find themselves in the false vacuum, and
consequently less and less bubbles will be nucleated. Also, and although in
a realistic situation one would expect the nucleation rate to vary with the
amount of dissipation present in the system, we kept the nucleation rate
constant throughout the simulation series. Since we will be concerned only
about the number of created vortices per nucleated bubble, $n_d$, this will
be of no consequence to us. What is of concern to us in this case however is
the number of bubbles used in the simulations. We imposed periodic boundary
conditions in our two dimensional box and tried to keep the number of
bubbles constant throughout simulations with different values of the
friction coefficient. For simulations with low $\gamma $, an average around
35 bubbles per run turned out to be sufficient. Figure 4 shows a particular
phase transition for $\gamma =2.5$ at four different stages. Space and time
are measured in units of the bubble wall thickness. Note how the
vortex-antivortex pair that appears close to $x=0,y=50$ at $t=550$ has
annihilated by the end of the transition, at $t=800$. This brings about the
question of exactly what vortices are we counting, or rather, of {\em when}
are we counting the vortices, since the number of vortices in the simulation
volume is a function of time that in our case will tend to zero as $%
t\rightarrow \infty $. As it is only the final state of the system after the
transition that we are interested in, we decided to count only those
vortices surviving when the transition was about 95\% completed. That is, in
this case, the six defects that appear in Fig.4(d) --three vortices and
three antivortices. However this rule of thumb has to be applied carefully
--in a sense that is made clear in Figs.5 and 6. In Fig.5 we have four shots
of a phase transition with $\gamma =10$, for $t=600,t=900,t=1200,$and $t=1500
$ --note the obvious difference in the time scale needed to complete the
transition. By $t=1500$ it would appear that we have two pockets of false
vacuum left in the central region, but which do not produce any winding of
the phase, then two vortices and perhaps and antivortex still closing at $%
x=0,y=80$ $.$ This would clearly violate the charge conservation that has to
be fulfilled in our torus however. The solution to the problem comes from
realizing that this last pocket of false vacuum actually has a winding
number $-2$, and therefore it is bound to decay into two antivortices that
will immediately start separating from each other. Figs. 6 (a) and (b) show
precisely this process as it takes place, blowing out the corresponding
region of simulation space. Note however that the process is completed only
well after the transition has finished. A ``correct'' counting of the
defects created in this case however would yield a value of 4. Note also how
in Fig. 5 the phase has a much smoother structure than in Fig.4. This is
easy to understand in physical terms: since we have periodic boundary
conditions, in the absence of dissipation all the false vacuum energy would
go into phase gradients and field oscillations. The higher the value that $%
\gamma $ takes however the more energy we dissipate and is not available to
form gradients. In Fig.7 finally we show the last stage of a transition with
a high value of $\gamma ,\gamma =75$. For such high values of $\gamma $ we
had to resort to larger simulations to obtain a significant number of vortex
creation events per run. Roughly, we multiplied by 4 the length of the box
side, therefore multiplying by 16 both the area and the average number of
nucleated bubbles --up to a total of about 570 bubbles per run. Note how,
although gradients of the phase of course persist in larger scales, at the
scale of Figs.4 and 5 the phase is even smoother now. We still have numerous
small pockets of false vacuum left that will need some extra time to close
completely. Most of them do not carry winding of the phase, but some of them
do however, such as the vortex at $x=100,y=25.$

For simulations with relatively low $\gamma $, that is, for those with about
35 bubbles per run, we performed 50 runs for each value of $\gamma $ in
order to get a good statistic for the average number of vortices created per
nucleated bubble . We computed $n_d$ separately for each run dividing the
total number of vortices existing at the end of the transition by the total
number of bubbles nucleated in that run. For those values of $\gamma $ that
required larger simulations however we could only perform 10 different
simulations. The total computer time involved in the project is estimated to
be around 500-600 hours in an Alpha 2100 DEC station. The results are shown
in Fig.8., where $n_d$ is plotted versus $\gamma $ in a loglog plot together
with a straight line fit. The first point corresponds to $\gamma =0.25$,
that is, a very nearly undamped regime. The value of $n_d$, $0.24\pm 0.01$,
that we get for this case agrees with the value of the number of nucleated
defects per bubble in the undamped case for 2 spatial dimensions quoted in
the literature \cite{sri92}, as could be expected. After all, if the
terminal velocity of the bubble walls is only slightly smaller than 1, then
the phase waves will need a large time interval in order to catch up with
the bubble walls, maybe even larger than the total time needed to complete
the transition, and so, we would still effectively be in the undamped
scenario. After that we see a rather flat plateau up until the next value of 
$\gamma ,\gamma =2.5$, from which point on friction will start to have a
noticeable effect. Starting at that point the data seem to suggest a soft
power law decay of $n_d$. We have plotted the least squares straight line
that fits that data from $\gamma =2.5$ to $\gamma =100$ with a continuous
line in Fig.8. This line has a slope of $-0.58\pm 0.05$, and thus we would
have $n_d\sim \gamma ^{-0.58\pm 0.05}$ with a 95\% confidence margin,
assuming of course that a simple regression is correct. The correlation
coefficient for this regression is of $-0.996$, which would seem to indicate
a good fit. However, it is still unclear at this point whether this is the
best we can do or whether the data admit a different interpretation. A
detailed analysis and interpretation of the data will be left to a following
paper in which the author is currently working jointly with J. Borrill.
Notice that in any case the decay is rather soft, especially if we take into
account that we are more likely to nucleate a larger number of bubbles in
slower transitions.

\section{Conclusions}

In first order phase transitions where a frictional coupling between the
scalar driving the transition and the rest of the matter content is
important, the mechanism for topological defects formation differs in some
important features from the one usually understood to take place in undamped
transitions. After having understood the detailed dynamics of two and three
bubble collisions and defect formation in a previous paper, the aim of this
paper was to try to quantify the effect that these differences have on the
probability of defect formation per bubble, $n_d$. In order to do that we
have simulated phase transitions for a set of different values of $\gamma $
ranging from the almost undamped case to $\gamma =100$, thus finding the
dependance of $n_d$ on $\gamma $ (i.e., on the terminal velocity for bubble
wall expansion, since $v_t\sim \gamma ^{-1}$). The total computer time
involved in the project is estimated to be around 500-600 hours on an Alpha
2100 DEC station. The main result of the paper are Figs.8 and 9, where a
soft power law decay of $n_d$ with $\gamma $ is shown.

\section{Acknowledgments}

This project was greatly enriched by discussions with Alex Vilenkin to whom
I wish to express my gratitude. Special thanks are also due to Alejandra
Melfo, for allowing me to use and enrich the code that she first developed.

\samepage

\end{document}